\documentstyle[prd,aps,epsfig,floats,tighten,preprint]{revtex}
\draft

\newcommand{\nature}[3]{Nature {\bf B#1}, #3 (#2)}

\renewcommand{\apj}[3]{Astrophys.\ J.\ {\bf #1}, #3 (#2)}
\renewcommand{\prl}[3]{Phys.\ Rev.\ Lett. {\bf #1}, #3 (#2)}
\renewcommand{\prd}[3]{Phys.\ Rev.\ {\bf D#1}, #3 (#2)}

\begin{document}


\title{
\null
\vskip-6pt \hfill {\rm\small MCTP-02-02} \\
\vskip-9pt~\\
Cardassian Expansion: a Model in which the Universe is Flat,
Matter Dominated, and Accelerating}

\vspace{.5in}

\author{Katherine Freese and Matthew Lewis}

\address{
\vspace{.7cm} Michigan Center for Theoretical Physics, University of
Michigan, Ann Arbor, MI 48109, USA\\}

\maketitle

\begin{abstract}

A modification to the Friedmann Robertson Walker equation is proposed in
which the universe is flat, matter dominated, and accelerating.  An
additional term, which contains only matter or radiation (no vacuum
contribution), becomes the dominant driver of expansion at a late epoch
of the universe. During the epoch when the new term dominates, the
universe accelerates; we call this period of acceleration the Cardassian
era.  The universe can be flat and yet consist of only matter and
radiation, and still be compatible with observations.  The energy
density required to close the universe is much smaller than in a
standard cosmology, so that matter can be sufficient to provide a flat
geometry.  The new term required may arise, e.g., as a consequence of
our observable universe living as a 3-dimensional brane in a higher
dimensional universe.  The Cardassian model survives several
observational tests, including the cosmic background radiation, the age
of the universe, the cluster baryon fraction, and structure formation.

\end{abstract}


Recent observations of Type IA Supernovae \cite{SN1,SN2} as well as
concordance with other observations (including the microwave background
and galaxy power spectra) indicate that the universe is accelerating.
Many authors have explored a cosmological constant, a decaying vacuum
energy \cite{fafm,frieman}, and quintessence
\cite{stein,caldwell,huey} as possible explanations for such
an acceleration.

Here we propose an alternative which invokes no vacuum energy whatsoever.
In our model the universe is flat and yet consists only of matter
and radiation.
Pure matter (or radiation) alone can drive an accelerated
expansion if the Friedmann Robertson Walker (FRW) equation is modified
by the addition of a new term on the right hand side as follows:
\begin{equation}
\label{eq:new}
H^2 = A \rho + B \rho^n ,
\end{equation}
where $H = \dot R / R$ is the Hubble constant (as a function of time),
$R$ is the scale factor of the universe, the energy
density $\rho$ contains only ordinary matter and radiation,
and we will take
\begin{equation}
n< 2/3.
\end{equation}
In the usual FRW equation
$B=0$.  To be consistent with the usual FRW result, we take $A={8\pi
\over 3 m_{pl}^2}$.  We note here that the geometry is flat,
as required by measurements of the cosmic background radiation
\cite{boom}, so that there are no curvature terms in the equation.
There is no vacuum term in the equation.
This paper does not address the cosmological constant ($\Lambda$) problem;
we simply set $\Lambda=0$.

In this paper, we first study the phenomenology of the ansatz in 
Eq.(\ref{eq:new}), and then turn to a discussion of the origin of
this equation\footnote{As discussed below, we were 
motivated to study an equation of this form
by work of Chung and Freese \cite{cf} who showed that terms
of the form $\rho^n$ can generically appear in the FRW equation
as a consequence of embedding our observable universe as a brane
in extra dimensions.}.  Directions for a future search for a fundamental
theory will be discussed.

{\it The Role of the Cardassian Term\footnote{The name Cardassian
refers to a humanoid race in Star Trek whose goal is to take over the universe,
i.e., accelerated expansion.  This race looks foreign to us
and yet is made entirely of matter.}:}
The new term in the equation (the second term on the right
hand side) is initially negligible.  It only
comes to dominate recently, at the redshift $z_{eq} \sim O(1)$ indicated
by the supernovae observations.  Once the second term dominates,
it causes the universe to accelerate.  We can consider the contribution
of ordinary matter, with
\begin{equation}
\label{eq:matter}
\rho = \rho_0 (R/R_0)^{-3}
\end{equation}
to this
new term.  Once the new term dominates the right hand side of the equation,
we have accelerated expansion.  When the new term is so large
that the ordinary first term can be neglected, we find
\begin{equation}
R \propto t^{2 \over 3n}
\end{equation}
so that the expansion is superluminal (accelerated) for $n<2/3$.
As examples, for $n=2/3$ we have $R \sim t$;
for $n=1/3$ we have $R \sim t^2$; and for $n=1/6$ we have $R \sim t^4$.
The case of $n=2/3$ produces a term in the FRW
equation $H^2 \propto R^{-2}$; such a term looks similar to a curvature
term but is generated here by matter in a universe with a flat geometry.
Note that for $n=1/3$ the acceleration is constant, for $n>1/3$ the
acceleration is diminishing in time, while for $n<1/3$ the acceleration
is increasing (the cosmic jerk).

The second term starts to dominate at a redshift $z_{eq}$ when $A
\rho(z_{eq}) = B \rho^n(z_{eq})$, i.e., when

\begin{equation}
\label{eq:one}
B/A = \rho_0^{1-n} (1+z_{eq})^{3(1-n)} .
\end{equation}
From evaluating Eq.(\ref{eq:new}) today, we have
\begin{equation}
\label{hubbletoday}
H_0^2 = A \rho_0 + B \rho_0^n
\end{equation}
so that
\begin{equation}
\label{eq:two}
A = H_0^2/\rho_0 - B \rho_0^{n-1} .
\end{equation}

From Eqs.(\ref{eq:one}) and (\ref{eq:two}), we have
\begin{equation}
\label{eq:B}
B = {H_0^2 (1 + z_{eq})^{3(1-n)} \over \rho_0^n [1+ (1+z_{eq})^{3(1-n)}]} .
\end{equation}
We have two parameters in the model: $B$ and $n$, or, equivalently,
$z_{eq}$ and $n$.  
Note that $B$ here is chosen to make the second term kick in
at the right time to explain the observations.  As yet we have
no explanation of
the coincidence problem; i.e., we have no explanation
for the timing of $z_{eq}$.  Such an explanation would
arise if we had a reason for the required mass scale of $B$.
The parameter $B$ has units of  mass$^{2-4n}$.  Later, we will
discuss the origin of the Cardassian term in terms of extra dimensions,
and discuss the origin of the mass scale of $B$.
As discussed below, to match the CMB and supernovae data we take
$0.3 \leq z_{eq} \leq 1$, but this
value can easily be refined to better fit upcoming observations.

{\it What is the Current Energy Density of the Universe?}

Observations of the cosmic background radiation show that the geometry
of the universe is flat with $\Omega_0=1$.  
In the Cardassian model we need to revisit
the question of what value of energy density today,
$\rho_0$, corresponds to a flat geometry.
We will show that the energy density required to close
the universe is much smaller than in a standard cosmology,
so that matter can be sufficient to provide
a flat geometry.

The energy density $\rho_0$ that satisfies Eq.(\ref{hubbletoday})
is, by definition, the critical density.
From Eqs.(\ref{eq:new}) and (\ref{eq:one}), we can write
\begin{equation}
H^2 = A [\rho + \rho_0^{1-n} (1+z_{eq})^{3(1-n)}\rho^n] .
\end{equation}
Evaluating this equation today with $A=8\pi/(3m_{pl}^2)$, we have
\begin{equation}
H_0^2 = {8 \pi \over 3 m_{pl}^2} \rho_0 [1+(1+z_{eq})^{3(1-n)}] .
\end{equation}
Defining $\rho_0 = \Omega_0 \rho_c$
we find that the critical density $\rho_c$ has been modified from
its usual value, i.e., the number has changed.
We find
\begin{equation}
\rho_c = {3 H_0^2 m_{pl}^2 \over 8 \pi [1+(1+z_{eq})^{3(1-n)}]} .
\end{equation}
Thus
\begin{equation}
\rho_c = \rho_{c,old} \times F(n)
\end{equation}
where
\begin{equation}
\label{eq:F}
F(n) = [1+(1+z_{eq})^{3(1-n)}]^{-1}
\end{equation}
and
\begin{equation}
\rho_{c,old} = 1.88 \times 10^{-29} h_0^2 {\rm gm/cm^{-3}}
\end{equation}
and $h_0$ is the Hubble constant today in units of 100 km/s/Mpc.
In Figure 1, we have plotted the new critical density $\rho_c$
as a function of the two parameters $n$ and $z_{eq}$.
For example, if we take $z_{eq}=1$, we find

\begin{equation}
F = (1/3,1/5,0.15)\,\,\, {\rm for} \,\,\,n=(2/3,1/3, 1/6)\,\,\,
{\rm respectively}.
\end{equation}
We see that the value of the critical density can be much lower
than previously estimated.
Since $\Omega_0=1$, we have today's energy density as
$\rho_0=\rho_c$ as given above \footnote{An alternate possible definition
would be to keep the standard value of $\rho_c$ and discuss the
contribution to it from the two terms on the right hand side
of the modified FRW equation.  Then there would be a contribution
to $\Omega$ from the $\rho$ term and another contribution from
the $\rho^n$ term, with the two terms adding to 1.  This is the
approach taken when one discusses a cosmological constant in lieu
of our second term. However, the situation here is different in
that we have only matter in the equation.  The disadvantage of this
second choice of definitions would be that a value of the energy density today
equal to $\rho_c$ according to this second definition would not correspond
to a flat geometry.},
i.e.,
\begin{equation}
\rho_0 = (1/3,1/5,0.15) \times
 1.88 \times 10^{-29} h_0^2 \,\,{\rm gm /cm}^{3} \,\,\, {\rm for} \,\,\,
n=(2/3,1/3,1/6)
\,\,\, \mbox{ and } z_{eq}=1 .
\end{equation}

For larger values of $z_{eq}$, the modification to the value of $\rho_c$
can be even larger.  Note the amusing result that for $z_{eq}=2$ and
$n=1/12$, we have $\rho_c = 0.046 \rho_{c,old}$ so that baryons would
close the universe (not a universe we advocate).

\begin{figure}[ht]
\centering
\includegraphics{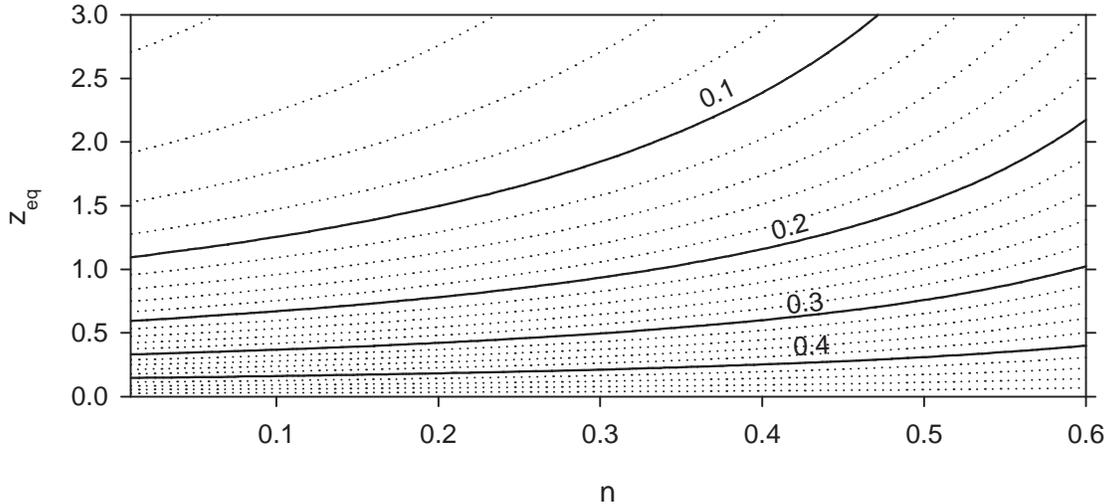}
\caption{The ratio $F(n,z_{eq})= \rho_c/\rho_{c,old}$ as given by
Eq.(\ref{eq:F}). The contour labeled 0.3 corresponds to parameters $n$
and $z_{eq}$ roughly consistent with present observations. }
\end{figure}

\bigskip
{\it Cluster Baryon Fraction}

For the past ten years, a multitude of observations has pointed
towards a value of the matter density $\rho_o \sim 0.3 \rho_{c,old}$. 
The cluster baryon fraction \cite{white,evrard} 
as well as the observed galaxy power spectrum are best fit
if the matter density is 0.3 of the old critical density.
Recent results from the CMB \cite{boom,dasi} also obtain this value.
In the standard cosmology this result implied that matter could
not provide the entire closure density.  Here, on the other hand,
 the value of the critical density can be much lower
than previously estimated.
Hence the cluster motivated value for $\rho_o$ is now
{\it compatible} with a closure
density of matter, $\Omega_o =1$, all in the form of matter.
For example, if $n= 0.6$ with $z_{eq} =1$,
or if $n=0.2$ with $z_{eq} = 0.4$, a critical density of matter
corresponds to  $\rho_o \sim 0.3 \rho_{c,old}$, as required
by the cluster baryon fraction and other data. In Figure 1, one can see which
combination of values of $n$ and $z_{eq}$ produce the
required factor of 0.3. 
If we assume that the value $\rho_o = 0.3
\rho_{c,old}$ is correct, for a given value of $n$
(that is constant in time) we can compute the value of $z_{eq}$ for our
model from Eq.(\ref{eq:F}).  Table I lists these values of $n$ and
$z_{eq}$.  Henceforth, we shall focus on these combinations
of parameters.

{\it Age of the Universe}

In the Cardassian model, the universe is older due to the presence of
the second term.   In Table I, we present the age of the universe
for various values of $n$ (under the assumption that
$\rho_0=0.3 \rho_{c,old}$).

\begin{table}
\begin{center}
\begin{tabular} {ccc}
$n$ & $z_{eq}$ & $H_0 t_0$ \\ \hline
0.60 & 1.00 & 0.73 \\
0.50 & 0.76 & 0.78 \\
0.40 & 0.60 & 0.83 \\
0.30 & 0.50 & 0.87 \\
0.20 & 0.42 & 0.92 \\
0.10 & 0.37 & 0.95 \\
0.00 & 0.33 & 0.99 \\
\end{tabular}
\end{center}
\caption{Values of ${z_{eq}}$ for various values of $n$ corresponding
to a universe with $\rho_0 = 0.3\rho_{c,old}$.  The age of the
universe today $t_0$ corresponding to the two parameters $n$ and
$z_{eq}$ is listed in the last column, where $H_0$ is the value of the
Hubble constant today.  }
\end{table}

If one takes $t_0 > 10$Gyr as the lower bound on globular cluster ages,
then one requires $t_0 H_0 > 0.66$ for $h_0=0.65$.
If one requires globular cluster ages greater than 11 Gyr
\cite{kc}, then $t_0 H_0 > 0.73$ for $h_0=0.65$.
All values in Table I satisfy these bounds.

{\it Structure Formation}

Since the new (Cardassian) term becomes important only at $z_{eq} \sim 1$,
early structure formation is not affected.  Below we discuss
the impact on late structure formation during the era
where the Cardassian term is important. This term accelerates
the expansion of the universe, and freezes out perturbation growth
once it dominates (much like when a curvature term dominates); this
freezeout happens late enough that it is relatively unimportant.
To obtain an idea of the type of effects that we may find, instead
of analyzing the exact perturbation equations with metric perturbations
included, we will merely modify the time dependence of the scale
factor in the usual Jeans analysis equation.
For now we take the standard equation for perturbation growth;
as a caveat, we warn that recent structure formation may be further
modified due to a change in Poisson's equations as described below.
For we now we take
\begin{equation}
\label{eq:struct}
\ddot{\delta} + 2(\dot R / R) \dot \delta = 4 \pi \rho \delta/m_{pl}^2
\end{equation}
where $\delta = (\rho - \bar \rho)/\bar\rho$ is the fluid overdensity.
Now one must substitute Eq.(\ref{eq:new}) for $\dot R/R$.
In the standard FRW cosmology with matter domination, $R \sim t^{2/3}$,
and there is one growing solution to $\delta$ with
$\delta \sim R \sim t^{2/3}$.  This standard result still applies
throughout most of the (matter dominated) history of the universe
in our new model, so that structure forms in the usual way.

Modifications set in once the new Cardassian term becomes important.
When $R \sim t^p$, Eq.(\ref{eq:struct}) can be written

\begin{equation}
\label{eq:struct2}
\delta ''(x) + {2 p \over x} \delta '(x) - {3 p^2 \over 2} x^{-3p}
\delta =0 ,
\end{equation}

\noindent where $x \equiv t/t_0$ with $t_0$ denoting the time today and
superscript prime refers to $d/dx$.  This equation can generally be
solved in terms of Bessel functions for constant $p$ (such as is the
case once the Cardassian term completely overrides the old term). A
simple example is the case of $n=2/3$ and $p=1$; in the limit $x>>3/4$,
the last term in Eq.(\ref{eq:struct2}) can be dropped and the equation
is solved as $\delta(t) = a_1 + a_2 t^{-1}$.  Perturbations cease
growing and become frozen in. This result agrees with the expectation
that in a universe that is expanding more rapidly, the overdensity will
grow more slowly with the scale factor.  As mentioned at the outset, as
long as the Cardassian term becomes important only very late in the
history of the universe, much of the structure we see will have already
formed and be unaffected.  Further comments on late structure formation
(e.g. cluster abundances) follow below.

{\it Doppler Peak in Cosmic Background Radiation}

Here we argue that the location of the first Doppler peak is only mildly
affected by the new Cardassian cosmology.  We need to calculate the
angle subtended by the sound horizon at recombination.  In the standard
FRW cosmology with flat geometry, this value corresponds to a spherical
harmonic with $l=200$. A peak at this angular scale has indeed been
confirmed \cite{boom}.  In the Cardassian cosmology we still have a flat
geometry. Hence, we can still write
\begin{equation}
\label{eq:theta}
\theta = s_*/d,
\end{equation}
 where $s_*$ is the sound horizon
at the time of recombination $t_r$ and ${d}$ is the distance
a light ray travels from recombination to today.  To calculate
these lengths, we use the fact that for a light ray $ds^2=0=
-dt^2 + a^2 d\vec{x}^2$ to write
\begin{equation}
\label{eq:light}
d = \int_{t_r}^{t_0} dt/a .
\end{equation}
Following the notation of Peebles \cite{peebles},
we define the redshift dependence of $H$ as
\begin{equation}
H(z) = H_0 E(z)
\end{equation}
so that Eq.(\ref{eq:light}) can be written
\begin{equation}
\label{eq:path}
d = {1 \over H_0 R_0} \int_{0}^{z_r} {dz \over E(z)} .
\end{equation}
Similarly, the sound horizon at recombination is
\begin{equation}
\label{eq:sound}
s_* = \int_{z_r}^\infty dt/a .
\end{equation}

In standard matter dominated FRW cosmology with $\Omega_{m,0} = 1$,
$E(z) = (1+z)^{3/2}$ in Eq.(\ref{eq:path}) and $d = 2/H_0 R_0$.

For the cosmology of Eq.(\ref{eq:new}), we have
\begin{equation}
\label{eq:newE}
E(z)^2 = F\times (1+z)^3 + (1-F) \times (1+z)^{3n}
\end{equation}

\noindent with $F$ given in Eq.(\ref{eq:F}).  As discussed previously,
as our standard case we will take $F \equiv \rho_c / \rho_{c,old} =
0.3$.  With this assumption, and by using expression Eq.(\ref{eq:newE})
in Eq.(\ref{eq:path}), we find that $d$ changes by a factor of (1.47,
1.88, 2.04, and 2.23) for $n$ =(0.6, 0.3, 0.2, and 0.1) respectively
compared to the the usual (nonCardassian) FRW universe with $\rho_{o} =
\rho_{c,old}$. In addition $s_* \propto {1 \over \sqrt{F}} {\rm
ln}{\sqrt{1+R_*} + \sqrt{R_* + r_*R_*} \over 1+ \sqrt{r_* R_*}}$ where
$r_* = 0.042 (F h^2)^{-1}$ and $R_* = 30 \Omega_b h^2$ and we use
$h=0.7$ and $\Omega_b=0.04$. We find that $s_*$ changes by a factor of
(1.44, 1.62, 1.67, 1.29) for $n$=(0.6, 0.3, 0.2, and 0.1) respectively
compared to the usual FRW universe with $\rho_{o} = \rho_{c,old}$.  The
angle subtended by the sound horizon on the surface of last scattering
decreases and the location ($l$) of the first Doppler peak increases by
roughly a factor of

\begin{equation}
\mbox{(1.02, 1.11, 1.12, 1.13)} \,\,\, {\rm for} \,\,\,
$n$ =\mbox{ (0.6, 0.3, 0.2, 0.1)} \,\,\, {\rm respectively}
\end{equation}

\noindent compared to the usual FRW universe with $\rho_{o} = \rho_{c,old}$. 
This shift still lies within
the experimental uncertainty on measurements of the location of the
Doppler peak.

We note the following: in the same way that a nonzero $\Lambda$ may make
the current CBR observations compatible with a small but nonzero
curvature, indeed a nonzero Cardassian term could also allow for a
nonzero curvature in the data.  A more accurate study of the effects of
Cardassian expansion on the cosmic background radiation (including the
first and higher peaks) is the subject of a future study.

{\it The Cutoff Energy Density}

An alternate way to write Eq.(\ref{eq:new}) is

\begin{equation}
\label{eq:alternate}
H^2 = A \rho \bigl[1+({\rho / \rho_{cutoff}})^{n-1}\bigr]
\end{equation}

\noindent where $\rho_{cutoff} \equiv \rho(z_{eq})= A/B$.  This notation
offers a new interpretation; it indicates that the second term only
becomes important once the energy density of the universe drops below
$\rho_{cutoff}$, which has a value a few times the critical
density. Hence, regions of the universe where the density exceeds this
cutoff density will not experience effects associated with the
Cardassian term.  In particular, we can be reassured that the new term
won't affect gravity on the Earth or the Solar System.  The density of
water on the Earth is 1 gm/cm$^3$, which is 28 orders of magnitude
higher than the critical density.

{\it Comparing to Quintessence}

We note that, with regard to observational tests, one can make a
correspondence between the Cardassian and Quintessence models; we
stress, however, that the two models are entirely
different. Quintessence requires a dark energy component with a
specific equation of state ($p = w\rho$), whereas the only ingredients
in the Cardassian model are ordinary matter ($p = 0$) and radiation
($p = 1/3$). However, as far as any observation that involves only
$R(t)$, or equivalently $H(z)$, the two models predict the same
effects on the observation.  Regarding such observations, we can make
the following identifications between the Cardassian and quintessence
models: $n \Rightarrow w+1$, $F\Rightarrow \Omega_m$, and $1-F
\Rightarrow \Omega_Q$, where $w$ is the quintessence equation of state
parameter, $\Omega_m= \rho_m/\rho_{c,old}$ is the ratio of matter
density to the (old) critical density in the standard FRW cosmology
appropriate to quintessence, $\Omega_Q= \rho_Q/\rho_{c,old}$ is the
ratio of quintessence energy density to the (old) critical density,
and F is given by Eq.(\ref{eq:F}).  In this way, the Cardassian model
can make contact with quintessence with regard to observational tests.

All observational constraints on quintessence that depend only on the
scale factor, $R(t)$ (or, equivalently, $H(z)$) can also be used to
constrain the Cardassian model.  However, because the Cardassian model
requires modified Einstein equations (see below), the gravitational
Poisson's equations and consequently late-time structure formation may
be changed; e.g., the redshift dependence of cluster abundance should
be different in the two models.  These effects (and others, such as
the fact that quintessence clumps) may serve to distinguish the
Cardassian and quintessence models. The correspondence with
quintessence, as well as discussion of distinguishing tests will be
the subject of a future paper.

{\it Best Fit of Parameters to Current Data}

We can find the best fit of the Cardassian parameters $n$ and $z_{eq}$
to current CMB and Supernova data.  The current best fit is obtained for
$\rho_o = 0.3 \rho_{c,old}$ (as we have discussed above) and $n<0.4$
(equivalently, $w < -0.6$) \cite{bean,hm}.  
In Table I one can see the values of
$z_{eq}$ compatible with this bound, as well as the resultant age of the
universe.  
As an example, for $n= 0.2$ (equivalently, $w=-0.8$), 
we find that $z_{eq} = 0.42$.
Then the position of the first Doppler peak is shifted by a factor of
1.12.  The age of the universe is 13 Gyr.  The cutoff energy density is
$\rho_{cutoff} = 2.7 \rho_c$, so that the new term is important only for
$\rho < \rho_{cutoff} = 2.7\rho_c$.  Hence, as mentioned above,
the Cardassian term won't affect the physics of the Earth or solar
system in any way.

We note the enormous uncertainty in the current data; future
experiments (such as SNAP \cite{snap}) will constrain these parameters
further.

{\it Extra Dimensions}

A Cardassian term may arise as a consequence of embedding our
observable universe as a 3+1 dimensional brane in extra dimensions.
Chung and Freese \cite{cf} showed that, in a 5-dimensional universe
with metric
\begin{equation}
\label{eq:metric}
ds^2 = -q^2(\tau,u) d\tau^2 + a^2(\tau,u) d{\vec{x}}^2 + b^2(\tau,u)du^2 ,
\end{equation}
where $u$ is the coordinate of the fifth dimension,
one may obtain a modified FRW equation on our observable
brane with $H^2 \sim \rho^n$ for any $n$ (see also \cite{binetruy}).
This result was obtained
with 5-dimensional Einstein equations plus the Israel boundary
conditions relating the energy-momentum on our brane to the
derivatives of the metric in the bulk.

We do not yet have a fundamental higher dimensional theory, i.e.,
a higher dimensional $T_{\mu\nu}$,
which we believe describes our universe.
Once we have this, we can write down the modified four-dimensional
Einstein's equations and compute the modified Poisson's equations,
as would be required, e.g., to fully understand latetime structure formation.

There is no unique 5-dimensional energy momentum tensor
$T_{\mu\nu}$ that gives rise to Eq.(\ref{eq:new}) on our brane.
Hence, in this paper we construct an example which is easy to find
but is clearly not our universe, simply as a proof that such
an example can be written down.
Following \cite{cf} (see Eqs. (24) and (25) there with $F(u) = u$),
we have constructed an example of a bulk
$T_{\mu\nu}$ for arbitrary $n$ in $H^2 \sim \rho^n$, matter
on the brane as in Eq.(\ref{eq:matter}), and with $q = b$ in
Eq.(\ref{eq:metric}). We display only $T_0^0$ here (the
other components will be published in a future paper):
\begin{equation}
\label{eq:t00}
{\kappa_5^2 T_0^0 = -\frac{ 3^{-\frac{4+n}{n}} B^{-\frac{2}{n}}\epsilon 
}{n^2\tau^2}
\left[ 4 \cdot 81^{\frac{1}{n}} B^{\frac{2}{n}} - 16^{\frac{1}{n}} \kappa^4_5
(\frac{1}{n\tau})^{\frac{4}{n}} (n^2\tau^2+u^2)\right]}
\end{equation}
where
\begin{equation}
{\epsilon =
\exp\left[-(2/3)^{\frac{2+n}{n}}B^{-\frac{1}{n}}\kappa^2_5
(\frac{1}{n\tau})^{\frac{2}{n}}u\right]}
\end{equation}
and the constant $\kappa_5$ is related to the 5-dimensional Newton's
constant $G_5$ and 5-D reduced Planck mass $M_5$ by the relation
$\kappa_5^2 = 8 \pi G_5 = M_5^{-3}$.
This is merely one (inelegant) example of many bulk $T_{\mu\nu}$ that
produce Cardassian expansion.

We may now investigate the meaning of the values of $B(n)$
required by  Eq.(\ref{eq:B}), where $B(n)$ is the
parameter in front of the new Cardassian term in Eq.(\ref{eq:new}).
As mentioned previously, the mass scale
of $B$ has units of $m^{2-4n}$. We find that the corresponding
mass scale is very small for $n<1/2$, is singular at $n=1/2$, and
then goes over to a very large value for $n>1/2$.  Specifically,
for $n=2/3$ and $z_{eq} = 1$,
we obtain $B \sim 10^{-52}$GeV$^{-2/3}$ which
corresponds to a mass scale of $10^{78}$GeV.  In the
context of extra dimensions, this large mass scale
turns out to cancel against other large numbers in such
a way that it corresponds to reasonable values of the energy
momentum tensor in the bulk.
We find that $\tau$ is roughly the
age of the universe and we have $\epsilon \sim 1$ for all $u$.
Then we have
\begin{equation}
\label{eq:tnumber}
T_0^0 \sim (10^{-5} {\rm GeV})^5.
\end{equation}
Although this value is not motivated, it is not unreasonable.
In other words, reasonable bulk values can generate the required
parameters in Eq.(\ref{eq:new}).
Numerical values for other components of $T_{\mu\nu}$ are the same
order of magnitude, with the exception that $T_{04} \sim 0$.
For the case of $n=1/3$, we obtain
a mass scale of $10^{-101}$, which cancels other small
numbers in such a way as to again require roughly Eq.(\ref{eq:tnumber})
to be satisfied.  The form of $T_{\mu\nu}$ given in Eq.(\ref{eq:t00})
is by no means unique and has been presented merely as an existence
proof; we hope a more elegant $T_{\mu\nu}$ may be found, perhaps
with a motivation for the required value of $B$.

{\it Discussion}

We have presented Eq.(\ref{eq:new}) as a modification to the FRW
equations in order to suggest an explanation of the recent acceleration
of the universe.  In the Cardassian model, the universe can be flat and
yet matter dominated.  We have found that the new Cardassian term can
dominate the expansion of the universe after $z_{eq} = \mathcal{O}$$(1)$
and can drive an acceleration.  We have found that matter alone can be
responsible for this behavior (but see the comments below).  The current
value of the energy density of the universe is then smaller than in the
standard model and yet is at the critical value for a flat geometry.
Structure formation is unaffected before $z_{eq}$.  The age of the
universe is somewhat longer.  The first Doppler peak of the cosmic
background radiation is shifted only slightly and remains consistent
with experimental results.  Such a modified FRW equation may result from
the existence of extra dimensions.  Further work is required to find a
simple fundamental theory responsible for Eq.({\ref{eq:new}).

Questions of interpretation remain.  We have said that
matter alone is responsible for the accelerated behavior.
However, if the Cardassian behavior results from integrating out
extra dimensions, then one may ask what behavior of the radii
of the extra dimensions is required. The Israel conditions
connect the energy density on the brane to fields in the bulk.
The required behavior of bulk fields is not transparent when
one writes the modified FRW equation.  We have found a large or small mass
scale to be required, which must result from
the extra dimensions. In principle one would like to have
a complete 5-dimensional theory so as to perform post-Newtonian
tests on the model and also to check other consequences.
For example, with a 5-dimensional model, one would like
to compare with limits from fifth force experiments and
to check that none of the higher dimensional fields are overcontributing
to the energy density of the universe at any point (the moduli problem).

One might attempt to use a Cardassian term
(the second term in Eq.(\ref{eq:new}))
to drive an early inflationary era in the universe
as well.  For $n<1/2$ one could have
a superluminal expansion during the radiation dominated era.
However, once accelerated expansion begins, Eq.(\ref{eq:new})
without a potential provides no way for inflation to stop.
Hence we have focused on using this new term to generate acceleration
today rather than to cause an inflationary mechanism early on.
However, it may be possible to combine such a term with a different way
to end inflation.

\acknowledgments

K.F. thanks Ted Baltz, Daniel Chung, Richard Easther, 
Gus Evrard, Paolo Gondolo, Wayne Hu, Lam Hui, Will Kinney,
Risa Wechsler, and especially Jim Liu for many useful conversations
and helpful suggestions.

We acknowledge support from the Department of Energy via the
University of Michigan.


\end{document}